\documentclass[12pt,preprint]{aastex}
\slugcomment{Accepted by the Astronomical Journal on 10 May 2012}
\shortauthors{Krist et al.}
\shorttitle{HD 202628}

\def\asec{$^{\prime\prime}$~}
\def\arcsec{$^{\prime\prime}$~}

\def\deg{$^{\circ}$}

\def\gapp{\lower 3pt\hbox{${\buildrel > \over \sim}$}\ }

\begin{document}

\title{\it Hubble Space Telescope Observations of the HD 202628 Debris Disk}

\author{John E. Krist\altaffilmark{1}, Karl R. Stapelfeldt\altaffilmark{2},
Geoffrey Bryden\altaffilmark{1}, Peter Plavchan\altaffilmark{3}}

\altaffiltext{1}{ Jet Propulsion Laboratory, California Institute of 
Technology, 4800 Oak Grove Drive, Pasadena CA 91109 }
\altaffiltext{2}{ Laboratory for Exoplanets and Stellar Astrophysics,
Code 667, NASA Goddard Space Flight Center, Greenbelt, MD 20771}
\altaffiltext{3}{ NASA Exoplanet Science Institute, California Institute 
of Technology, 770 S Wilson Ave, Pasadena, CA 91125 }

\begin{abstract}

A ring-shaped debris disk around the G2V star HD 202628 ($d$ = 24.4 pc) was
imaged in scattered light at visible wavelengths using the coronagraphic mode
of the Space Telescope Imaging Spectrograph on the {\it Hubble Space
Telescope}.  The ring is inclined by $\sim$64\deg\ from face-on, based on the
apparent major/minor axis ratio, with the major axis aligned along PA =
130\deg.  It has inner and outer radii ($>50$\% maximum surface brightness) of
139 AU and 193 AU in the northwest ansae and 161 AU and 223 AU in the southeast
($\Delta r/r \approx 0.4$). The maximum visible radial extent is $\sim254$ AU.
With a mean surface brightnesses of V $\approx$ 24 mag arcsec$^{-2}$, this is
the faintest debris disk observed to date in reflected light.  The center of
the ring appears offset from the star by $\sim$28 AU (deprojected). An ellipse
fit to the inner edge has an eccentricity of 0.18 and $a$ = 158 AU.  This
offset, along with the relatively sharp inner edge of the ring, suggests the
influence of a planetary-mass companion.  There is a strong similarity with the
debris ring around Fomalhaut, though HD 202628 is a more mature star with an
estimated age of about 2 Gyr. 

We also provide surface brightness limits for nine other stars in our study
with strong {\it Spitzer} excesses around which no debris disks were detected in
scattered light (HD 377, HD 7590, HD 38858, HD 45184, HD 73350, HD 135599, HD
145229, HD 187897, and HD 201219).
 
\end{abstract}

\keywords{circumstellar matter --  stars: individual (HD 202628, HD 377, HD 7590, 
HD 38858, HD 45184, HD 73350, HD 135599, HD 145229, HD 187897, HD 201219)}
\vfil\eject

\section{Introduction}

Circumstellar debris disks, clouds of dust created by collisions of
planetesimals such as comets and asteroids, provide evidence that stars are the
central hosts of dynamic systems.  Disk structures such as clearings, gaps, and
asymmetrical dust distributions may indirectly reveal the presence of planets.
The most prominent example is the Fomalhaut debris disk (Kalas et al.\ 2005)
and its apparent planetary-mass companion, Fomalhaut b (Kalas et al.\ 2008;
Chiang et al.\ 2009).  The presence of dust around a star is most easily
discerned by a measured infrared excess, though the disk is usually unresolved.
While still relatively rare, resolved images of nearby debris disks spanning
the millimeter to visible wavelength range have substantially increased in
recent years, especially with the use of the {\it Hubble}, {\it Spitzer}, and
{\it Herschel} space telescopes. The dust is often concentrated in a ring, as
with Fomalhaut (Kalas et al.\ 2005), HD 4796 (Schneider et al.\  2009), and HD
207129 (Krist et al.\ 2010).  This  makes detection easier because the surface
brightness is greater than it would be in a more extended disk (e.g., $\beta$
Pictoris).

In an effort to expand the number of resolved debris disks, which so far number
around 20, we undertook an imaging survey using the {\it Hubble Space
Telescope} ($HST$).  For a list of candidate stars with {\it Spitzer}-measured
infrared excesses and emission ratios of $L_{dust}/L_{star} > 10^{-4}$ we
predicted the detectability of a ring-shaped disk around each assuming a width
of $\Delta r/r = 0.2$, an albedo of 0.1, a canonical grain size distribution,
and fractional scattered light brightness based on the $L_{dust}/L_{star}$.
Ten nearby, solar-type targets with predicted inner radii outside of the inner
working angle of the $HST$ Space Telescope Imaging Spectrograph (STIS)
coronagraph were chosen.  Our goal was to combine the dust distribution seen in
the $HST$ images with {\it Spitzer} measurements to derive the actual grain
properties (spatial distribution, albedo, scattering phase function, grain size
distribution, etc.).  

Of the ten stars in our program, we have imaged a disk around only one, HD
202628 (HIP 105184, GJ 825.2, SAO 230622), a G2V star (V = 6.75) located at
24.4 pc.  Koerner et al.\ (2010) measured a $\lambda$ = 70 $\mu$m excess
($17\times$ the photosphere) using {\it Spitzer}, deriving $L_{dust} / L_{star}
= 1.4 \times 10^{-4}$.  A central clearing is indicated by the lack of any 24
$\mu$m excess that would be emitted by warmer dust near the star.  Using the
assumptions discussed above, we predicted an inner disk radius of 80 AU
(3\farcs 4).

\section{OBSERVATIONS AND DATA PROCESSING}

STIS observations of HD 202628 were taken on 15 May 2011 as part of {\it HST}
program 12291 (file prefixes obhs05 and obhs06). As with all of the targets in
this program, it was imaged over two consecutive orbits with the telescope
rolled by 28\deg\ between them.  Each orbit began with a coronagraphic target
acquisition image, and then the star was placed behind the occulting wedge
(WEDGEA1.8 aperture position, 0\farcs 9 mask radius).  Eight exposures of 282
sec each (2256 sec total per orbit) were taken.  The relatively short exposure
times were required to reduce saturation.  We initially used combined and
cosmic-ray-rejected images (\_crj.fits files) from the {\it HST} calibration
pipeline but found that the drift in the star position between exposures was
large enough to create significant artifacts after point spread function (PSF)
subtraction, since the drift is not accounted for by the pipeline.  We
therefore manually registered each exposure to the first image in the first
orbit to within 0.05 pixels (3 mas) using damped sinc interpolation.  This was
done by adjusting the shift of each image until the radial streaks from the
stellar PSF were visually minimized. We combined the results for each orbit
using our own cosmic ray rejection routine. 

Despite using the coronagraphic mode, scattered and unsuppressed diffracted
light from the star dominates the images, so PSF subtraction is required.
Taking advantage of the relative stability of {\it HST}, this is usually done
by observing another star of similar color and subtracting its scaled and
registered image.  STIS allows for only unfiltered imaging in its coronagraphic
mode, so the effective wavelength bandpass is set by the CCD detector and is
therefore quite broad (250 - 1100 nm).  Because the fine PSF structure is
dependent on object color, even minute color differences between the target and
reference PSFs can introduce substantial residual artifacts that might
overwhelm a faint disk.  We therefore chose to not observe a reference star and
instead used the iterative roll subtraction algorithm (Figure 1) that was successfully
applied to {\it HST} observations of the HD 207129 debris disk (Krist et al.\
2010).  This method dispenses with the need for another orbit to obtain a
reference PSF and avoids any color mismatch issues.  The main drawback is that
any disk seen in a face-on (or nearly so) orientation will essentially subtract
itself out.  Krist et al.\ demonstrated via modeling that this method produces
reliable results for inclined disks.  To set the roll pivot point, the position
of the star behind the mask was derived from the intercept of the diffraction
spikes; experiments indicate that the position is accurate to with 0.1 pixel (5
mas). 

As a comparison, we performed PSF subtraction using images of other stars from
our program, HD 45184 and HD 73350 (stellar properties are listed in Appendix
A).  These images were registered and scaled in intensity to match HD 202628
prior to subtraction.  The results using these reference PSFs (Figure 2) show
larger residuals, especially near the star, and worse definition. 

We note that there are dozens of small galaxies distributed over the field,
some barely resolved.  They introduce a significant source of confusion for
planet searches in this system.  A common proper motion determination of any
companion candidate will be a necessity. Fortunately, this star has a high
proper motion (241 mas yr$^{-1}$), so it would not take long to show that a
source is co-moving.  There are no obvious point sources inside the disk or
within 4\arcsec of the outer edge.

\section{RESULTS}

\subsection{General Disk Morphology}

As shown in Figure 1, positive and negative images of a disk, along with
background sources, are apparent when the image from the second roll is
directly subtracted from the one from the first.  Application of the iterative
subtraction algorithm reveals that the disk is a ring inclined from face-on by
$\sim$$64$\deg\ (based on the apparent major/minor axes ratio) with the
apparent major axis aligned along PA = 130\deg. Most of the disk along the
minor axis is obscured by large subtraction residuals, the STIS occulter, and
the $HST$ diffraction spikes.  Within $r <$ 3\farcs 5 (85 AU) there is a  halo
of unsubtracted starlight; such residuals are expected due to PSF mismatches
caused by time-dependent telescope aberrations and pointing errors (Krist
2004).

\begin{figure}
\epsscale{0.9}
\plotone{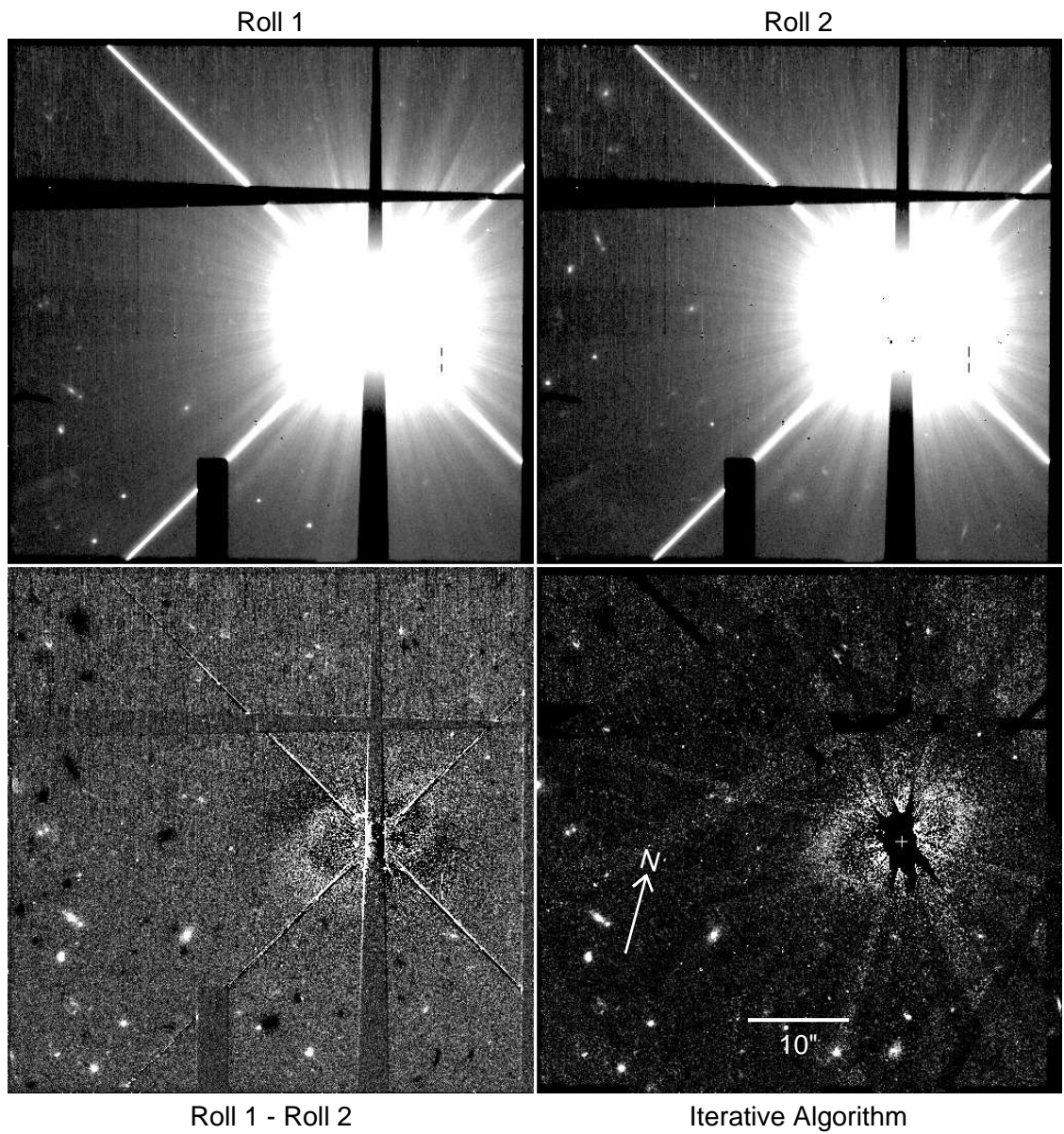}
\caption{$HST$ STIS coronagraphic observations of HD 202628 showing the full field of
the detector. The raw, PSF-unsubtracted 
images from the two telescope orientations, separated by 28\deg, are shown in
the top row, scaled between 0 -- 700 photons per pixel per 2256 sec exposure with a square root
intensity stretch. The shadows of the coronagraph's occulting masks are apparent.
The lower left panel shows the direct subtraction of the
second orientation image from the first, linearly scaled between -20 -- 70
photons per pixel per 2256 sec exposure. Note the positive and negative images of the disk at
different orientations.  The lower right panel shows the result from the iterative
roll subtraction algorithm, linearly scaled between 0 -- 70 photons per pixel
per 2256 sec.
The position of the star is marked by a cross.}
\epsscale{1.0}
\end{figure}

\begin{figure}
\epsscale{0.4}
\plotone{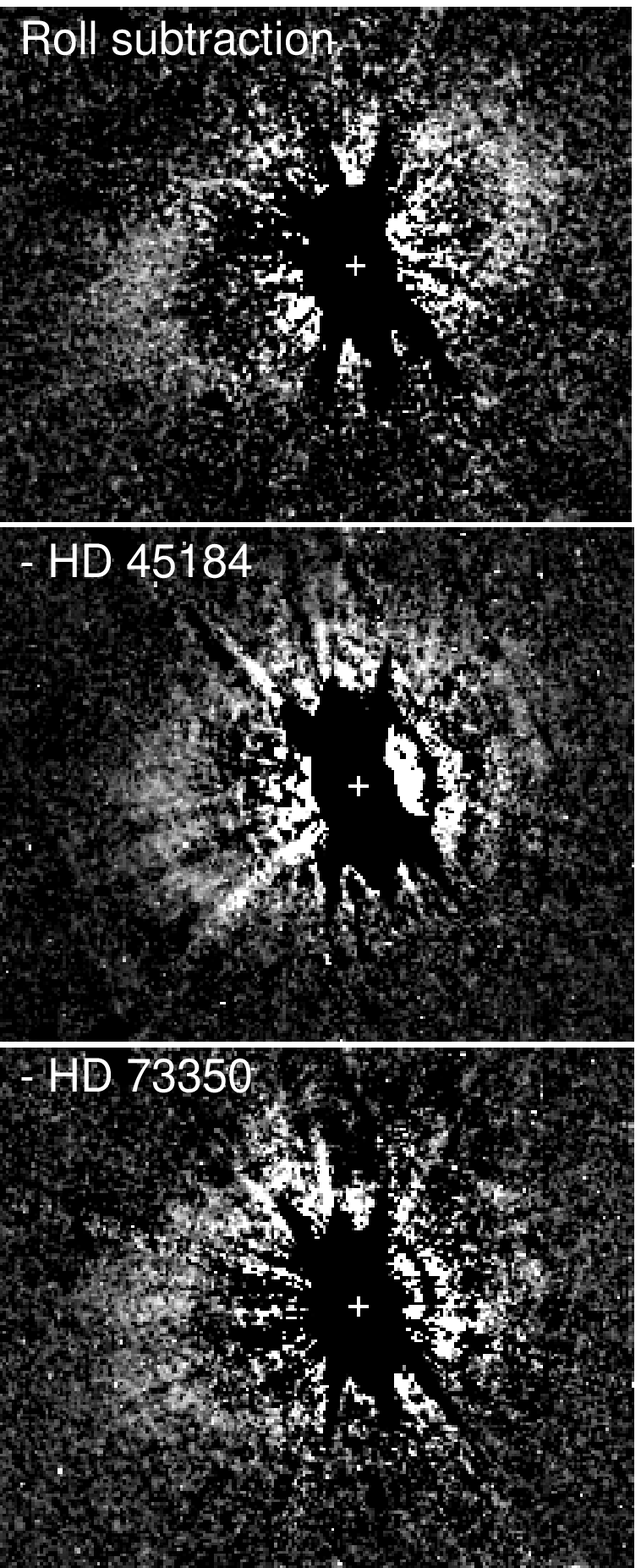}
\caption{STIS coronagraphic observation of HD 202628 with the stellar PSF
subtracted using (TOP) the iterative roll subtraction algorithm, (MIDDLE)
an image of HD 45184, and (BOTTOM) an image of HD 73350. Same orientation
as Figure 1. } 
\epsscale{1.0}
\end{figure}

The ring is asymmetric.  The inner and outer apparent edges of the northwest
ansa are at 5\farcs 8 (142 AU) and 8\farcs 7 (212 AU), respectively.  In the SE
ansa they are at 6\farcs 6 (162 AU) and 10\farcs 4 (254 AU).  The outer radius
is noise limited, so based on the apparent image the fractional width is
$\Delta r / r \approx 0.4$.

There are azimuthal brightness asymmetries, as shown in Figure 3. The ring
appears clumpy due to the noise, but it is probably azimuthally smooth.  The
per-pixel signal-to-noise ratio is about 1.0 in the ansae; the mean surface
brightness in the SE ansa is $\sim80\times$ less than the PSF at the same
location before subtraction.  We would not expect any significant amount of
material within the ring interior due to the lack of 24 $\mu$m emission
(Koerner et al.\ 2010).  Without any color information, we assumed neutral
scattering by the dust and measured peak and mean surface brightnesses,
respectively, of V = 23.6 and 24.0 mag arcsec$^{-2}$ ($\pm0.2$ mag) in the NW
ansa.  \footnote{Without a direct measurement of the star's brightness for
photometric calibration, we used the flux rate provided by the STIS Exposure
Time Calculator of $5.30 \times 10^{7}$ photons sec$^{-1}$ for a G2V V=6.75
star.} The SE ansa is about half as bright.  Visible portions of the NE section
of the ring are about 20\% -- 50\% brighter than the opposite side.  For
comparison, the Fomalhaut ring has a surface brightness of $\sim21.5$ mag
arcsec$^{-2}$ (Kalas et al.\ 2005) and HD 207129's is 23.7 (Krist et al.\
2010).

\begin{figure}
\epsscale{0.5}
\plotone{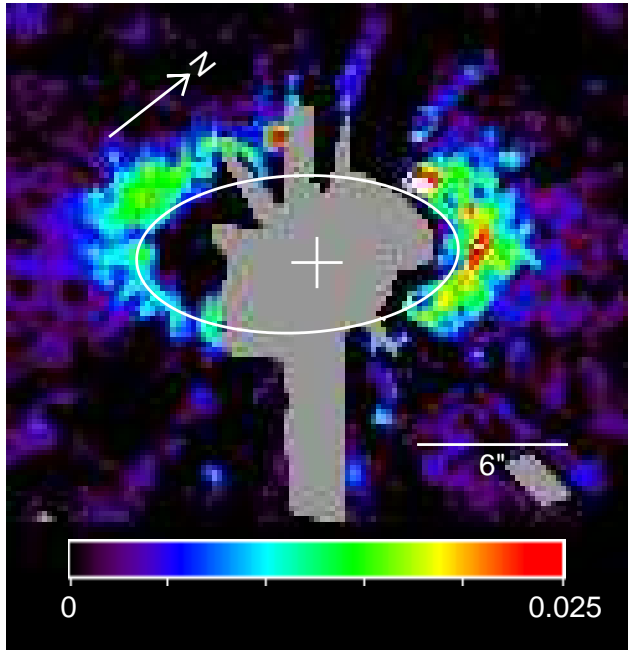}
\caption{Surface brightness map of the HD 202628 disk. The image has been
median filtered and rebinned to lower sampling to reduce noise.  The position
of the star is marked by a cross. An ellipse, fitted to the inner ring edge in
the original image,
is superposed. Known regions of high residuals or obscurations are colored
grey. The measured count rate (photons per second per 0\farcs 051 $\times$ 0\farcs 0.51 pixel) is shown in the 
linear intensity scale legend (0.025 phot sec$^{-1}$ pixel$^{-1}$ is equivalent to V = 23.6
mag arcsec$^{-2}$, assuming neutral scattering).  This figure is shown in color in the online publication.}   
\epsscale{1.0}
\end{figure}

\subsection{Ring Eccentricity}

The ring is eccentric, and the offset of the ring center from the star is
apparent in Figure 4.  Due to the low signal that prevented a reliable
least-squares fit, we visually fitted an inclined ellipse to the ring inner
edge while fixing one focus to the position of the star.  The size of the ring,
its aspect ratio, position angle, and two dimensional offset of the star from
center of the ellipse provides five independent observables.  These are
sufficient to constrain the five parameters needed for the model ellipse:
semi-major axis, eccentricity, inclination, major axis position angle, and
argument of periapse relative to the sky plane.  Trial model runs showed that
the semi-major axis, position angle, and inclination are strongly constrained,
while the eccentricity, and argument of periapse are slightly degenerate with
each other.  The latter interaction arises in this case because the stellar
offset from the ring center is aligned along its projected minor axis, rather
than along the major axis as in the case of Fomalhaut (Kalas et al. 2005).

The best fit parameters are an inner edge semi-major axis of 158 AU at
PA=134\deg, inclination of 61\deg, $e$ = 0.18, and ellipse major axis aligned
60\deg\ behind the plane of the sky.  While a smaller eccentricity combined with
a major axis close to the sky plane would adequately fit the observed image,
the parameters reported provided a much better fit to the deprojected image of
the system.  We estimate a $\pm0.02$ uncertainty in $e$. The geometric center
of the ellipse is offset from the star in the non-deprojected image by 19.5 AU
(16.0 AU along PA=130\deg\ and 11.3 AU along PA=40\deg). Deprojected, the
offset is 28.4 AU (16.0 AU along PA=130\deg and 23.4 AU along PA=40\deg).

\begin{figure} 
\epsscale{0.4} 
\plotone{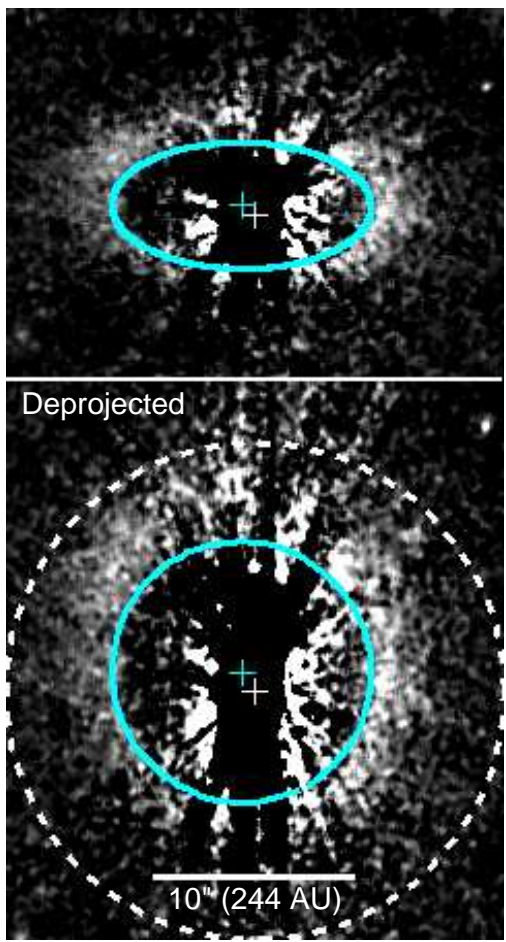} 
\caption{Direct (top) and deprojected (bottom) images of the HD 202628 disk,
shown with linear intensity stretches.  PA = 40\deg\ is towards the top. The
position of the star is marked with a white (lower right) cross.  An ellipse fitted to the
ring inner edge (same ellipse as is shown in Figure 3) is plotted as a solid blue line, and
the ellipse geometric center is marked with a blue (upper left) cross. A 300 AU radius dashed 
circle centered on the star is shown for comparison.  Interior to the ring
there are large PSF subtraction residuals, and the occulting mask obscures portions
of the ring.  Both images have been median filtered and $2\times2$ rebinned to
lower sampling to improve the signal.  This figure is shown in color in the
online publication.} 
\epsscale{1.0} \end{figure}

\subsection{{\it Spitzer} Observations} 

Koerner et al.\ (2010) observed HD 202628 with $Spitzer$/MIPS, finding a strong
70 $\mu$m excess but only a photospheric flux density at 24 $\mu$m.  An upper
limit to the 24 $\mu$m excess can be assumed to be 3$\times$ the uncertainty in
the absolute calibration, which is approximately 10\% of the 24 $\mu$m flux
density or 3.9 mJy.  This, in combination with the 70 $\mu$m excess of 100 mJy,
implies a dust temperature upper limit of $\sim$65 K.  If a minimum grain size
of a few microns and silicate emissivities are assumed (following Krist et al.
2010), the dust would need to be located at least 80 AU from the star to
reproduce the $Spitzer$ photometry.  The observed ring inner radius of 160 AU
implies a substantially colder dust temperature than the current $Spitzer$
limit.  Upcoming {\it Herschel} observations should clarify the spectral energy
distribution and characteristic dust temperature for the disk.

The ring's diameter of $\sim$14\asec (300 AU) observed with STIS is large
enough to produce a resolved 70 $\mu$m source to $Spitzer$/MIPS.  Koerner et
al.\ (2010) do not report any size information for their 70 $\mu$m detections,
so we investigated the source size in post-BCD mosaics retrieved from the
$Spitzer$ Heritage Archive.  A 2-D elliptical Gaussian fit to the 70 $\mu$m
source has a full width at half maximum of 22\farcs 1$\times$19\farcs 1
extended along position angle 138\deg.  Quadrature subtraction of the nominal
MIPS 70 $\mu$m 16\asec beam suggests an intrinsic source size of
15$^{\prime\prime}\times$10\asec, consistent with the inclined ring seen in
scattered light and elongated at essentially the same PA.  A small 0\farcs 6
offset between the centers of the 24 and 70 $\mu$m sources is not significant
given the 70 $\mu$m signal-to-noise ratio of 11 (Koerner et al.  2010), so an
assessment of a possible pericenter glow (Wyatt et al.\ 1999) awaits more
sensitive {\it Herschel} data.

\section{Discussion}

\subsection{Radial Profile}

Figure 5 shows the ring's radial (relative to the disk center) surface
brightness profile. It was derived from the filtered, deprojected image by
computing the median value at each radius within the two 90\deg\ sectors on
opposite sides of the star aligned along the apparent major axis of PA=130\deg\
(horizontal axis in the deprojection).  Note that no correction for azimuthal
brightness variations was applied.  The azimuthal root-mean-square values were
concurrently determined.  The radius was measured from the fitted ellipse's
geometric center.  The low signal-to-noise ratio of the image results in large
error bars, but a rapid dropoff along the inner edge is evident and indicates a
rather sharp truncation. However, we cannot discern whether the slope is due to
the difference between a mid-plane radial density variation, the vertical disk
structure seen integrated along the lines of sight, or a combination of both,
though for a geometrically-and-optically thin disk the projection effect would
be insignificant.  The profile extends above the field noise (which has an
RMS equivalent surface brightness of $\sim24.7$ mag arcsec$^{-2}$) over 120 -- 300 AU
($\Delta r / r \approx 0.7$). Between 150 -- 220 AU ($\Delta r / r = 0.4$) the
intensity is $>50$\% maximum brightness, which is commonly used to define
widths of ring-shaped disks.  Rough estimates of radial power laws that
describe the different zones of the surface brightness profile were determined:
$r^{12}$ (100 AU $< r <$ 156 AU), $r^{2}$ (156 AU $< r <$ 182 AU), $r^{-4.7}$
($r >$ 182 AU).  The true outer radius is indeterminate due to the noise.  

\begin{figure}
\epsscale{0.5}
\plotone{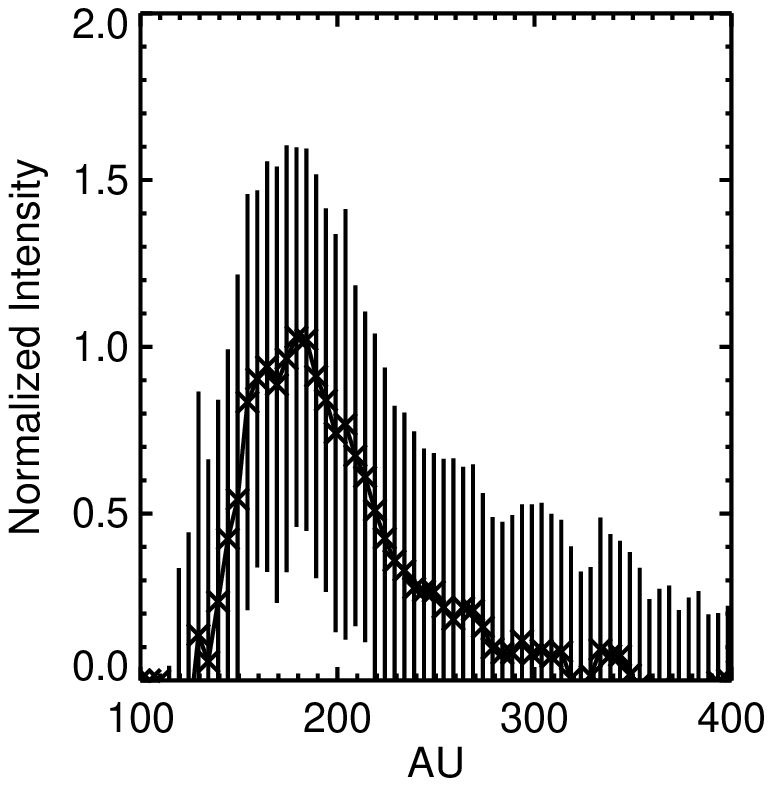}
\caption{Radial surface brightness profile of the HD 202628 ring
measured using the azimuthal median within 90\deg\ sectors centered on PA = 130\deg\ and 210\deg\ in the 
filtered, deprojected image.  The root-mean-square error bars are included. }
\epsscale{1.0}
\end{figure}

\subsection{Azimuthal Brightness Variations}

The western inner edge of the ring is closest to the star.  A line drawn
through the star and along PA=130\deg, the apparent major axis of the ring
(i.e., a horizontal line in Figure 4), intercepts the ring (as located by the
fitted ellipse) at distances 137 (NW) and 172 (SE) AU from the star.  In an
azimuthally uniform disk, the brightness at the SE intercept of the line and
ring should be 63\% of that in the NW due to the difference in distances from
the star.  The SE is actually more than 50\% fainter, so the ring probably does
not have an azimuthally uniform dust distribution.  Since both locations are at
the same scattering angle, the degree of forward scattering does not affect the
ratio.  However, if we assume isotropic scattering, then the SW side, being
closer to the star, should be brighter than the NE, but it is not.  If we
assume that the NE side of the ring is closest to us, forward scattering could
cause the NE/SW brightness asymmetry. 

To demonstrate this qualitatively and to constrain the degree of forward
scattering, we generated three-dimensional scattered light models.  For
simplicity, a circularly-symmetric, azimuthally-uniform disk with the measured
offset from the star was implemented (the deviation of an elliptical ring from
a circular one would not qualitatively change the results; the bulk of the
illumination effect is due to the ring offset from the star).  The model's dust
distribution was specified by the three power laws described previously (though
with +2 added to the exponents to account for the additional $r^{-2}$ stellar
illumination falloff in the surface brightness).  Without any strong constraint
from the noisy data, an outer disk radius of 450 AU was assumed.  We used a
flat (non-flaring) disk with a 2 AU scale height (this is largely unimportant
for these purposes).  The Henyey-Greenstein scattering phase function was used
with the amount of forward scattering defined by the parameter $g$ (0 =
isotropic, 1 = full forward scattering). To aid visual comparisons with the
data, the models were PSF convolved, and then Poisson and read noise was added.
Given the poor signal-to-noise level, we did not attempt to fit these models to
the data but used them only to provide qualitative comparisons. 

The models are shown in Figure 6.  The top side of the model ring is closest to
us.  With isotropic scattering ($g$ = 0), regions of the ring closest to the
star are brighter. Once forward scattering is introduced, the near side becomes
brighter, though the left/right asymmetry remains.  It appears from a
qualitative comparison with the data that the grains are only mildly forward
scattering ($0.15 < g < 0.25$).

\begin{figure}
\epsscale{1.0}
\plotone{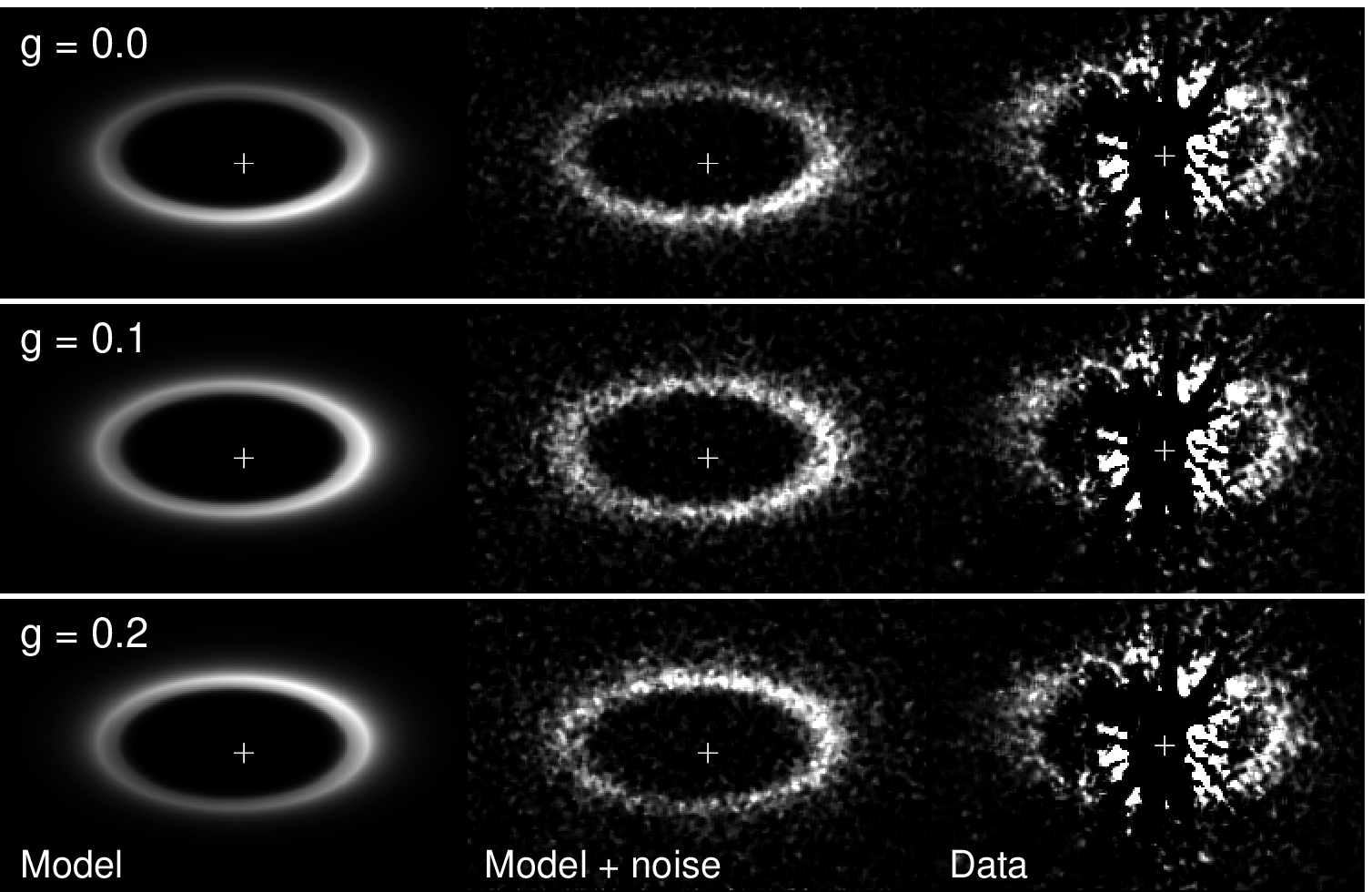}
\caption{(LEFT) Three-dimensional scattered light models of a circular disk,
including illumination from a star offset by the measured amount, with different degrees of forward
scattering described by the Henyey-Greenstein scattering asymmetry parameter $g$. 
The near side of the ring is towards the top. The lower right section of the
ring is closest to the star. When scattering is isotropic, the region close to
the star is brightest.  When forward scattering is introduced, the region closest
to the observer is brighter, though the left/right asymmetry due to the star
offset remains.  (MIDDLE) The disk models
with noise added. (RIGHT) The observed disk image for comparison (PA = 40\deg\ is
up).  All of the model and observed images are displayed as the square of the intensity ($I^2$)
to emphasize the brightness variations. The location of the star is marked with a cross. 
}
\epsscale{1.0}
\end{figure}

\subsection{The Age of HD 202628}

Valenti \& Fischer (2005) placed HD 202628 on a theoretical isochrone using its
distance and absolute luminosity (assumed to be solar) to derive a best-fit age
of 5.0 Gyr and an allowable range of 1.8 -- 6.8 Gyr.  Holmberg et al.\ (2009)
estimated 5.9 Gyr with an even wider range of 1.6 -- 10.2 Gyr, also based on
isochrone fitting.  However, the Ca II H and K lines suggest activity
consistent with a younger star (log~$R'_{\mathrm{HK}}$ = -4.73; Henry et al.
1996), as does the X-ray emission (log~$L_X$ = 28.41; H{\"u}nsch et al.\ 1999).
Using the methods described by Plavchan et al.\ (2009; hereafter P09), we
derive ages of 2.3 Gyr using $R'_{\mathrm{HK}}$ and 1.2 Gyr using $L_X$.  The
star is clearly younger than the Sun but older than 1 Gyr, so we assume an age
of $2.3 \pm 1$ Gyr.  This makes HD 202628 one of the oldest stars with an
imaged debris disk.

\subsection{Is the Ring Evidence of a Planet?}

The eccentricity and sharp inner edge of Fomalhaut's ring suggested
perturbations from a planetary mass object (Kalas et al. 2005).  Quillen (2006)
predicted that a planet should be orbiting just inside the ring edge and with
the same eccentricity.  The discovery of Fomalhaut b near this predicted
location appeared to validate this model (Kalas et al. 2008).  The similar ring
structures between HD 202628 and Fomalhaut likewise suggest the action of
planetary perturbations in this new system.  Two noteworthy differences are the
higher eccentricity and broader width (in $\Delta$r/r terms) of HD 202628's
ring.  The former likely reflects a higher eccentricity for the perturbing
planet.  A possible explanation for the latter is suggested by the Chiang et al.
(2009) models for the spatial distribution of small dust particles derived
from parent bodies in an eccentric ring.  Their Figure 3 shows how the
steepness of the ring inner edge can be produced by either a lower-mass planet
close to the ring or a higher mass planet at a greater interior distance.
Predictions for the outer profile differ substantially between these two
scenarios, however, depending on mass and orbital radius.  Specific simulations
for HD 202628 are needed, but a semi-major axis of $\sim$ 120 AU and mass $<10$
M$_{Jupiter}$ seem plausible. 

There are significant limits to detecting planets at large orbital radii ($>100$
AU) and older than a couple hundred million years in the infrared (due to low 
thermal emission from mature planets) and in the visible with future space 
coronagraphs (due to low reflected light contrast, even from giant planets).
Debris disk structures are thus critical, indirect indicators of planetary
system properties at large radii.  The structure of debris rings such as HD
202628 and Fomalhaut indicate that planets must exist at larger scales than
previously thought, a significant challenge for planet formation theory.  It is
believed that in-situ core accretion cannot take place rapidly enough at these
distances to form a planet before the dissipation of the primordial disk
(Dodson-Robinson et al.  2009).  In Fomalhaut, HR 8799 (Marois et al. 2008; Su
et al. 2009), and now HD 202628, the distant planets are seen or inferred
within their circumstellar debris belts - thus indicating they must have formed
from within a circumstellar disk and not (as binary stars do) from adjacent
collapsing cloud cores.  Both Fomalhaut and HR 8799 are A stars, and their
protoplanetary disks would likely have been fairly massive.  HD 202628, a G
star, would have had a less massive accretion disk, so the presence of a planet
even further from the star than those around the A stars would be a mystery.
The eccentricity of the purported planet, as suggested by that of the ring,
would indicate that it formed closer to the star and then scattered outward
by a more massive planet. Such a planet could not have been captured from
outside the system, otherwise the bodies creating the debris disk would have
been scattered.

\subsection{Grain Properties and Collision Timescales}

The deprojected image was used to estimate a disk/star scattered light
fraction.  The total flux was measured within the two 90\deg\ sectors centered
on PA=130\deg\ and 310\deg\ (along the ring apparent major axis) and the sum
doubled to account for the obscured sections of the ring along the line of
sight.  By this technique, $F_{scat} = F_{disk}/F_{star} = 6.8 \times 10^{-6}$
with an estimated error of $\pm8$\%.  We can make a rough estimate of the
albedo by computing the ratio of the scattered light fraction to the sum of
scattered and emitted flux fractions ($F_{emit} = L_{dust} / L_{star} = 1.4
\times 10^{-4}$ from Koerner et al.\ (2010)): $a \approx F_{scat} / (F_{scat} +
F_{emit}) \approx$ 4.6\%.  This is quite similar to HD 207129 ($F_{emit} =
L_{dust} / L_{star} = 7.6 \times 10^{-6}$, $a \approx $ 5\%; Krist et al.
2010).  Note that Fomalhaut has an intrinsically fainter disk ($F_{scat} = 9.6
\times 10^{-7}$), but the star is much brighter ($V$ = 1.2) so its disk has a
higher apparent surface brightness ($V \approx 21.5$ mag arcsec$^{-2}$).  

Grain-grain collisions dominate the evolution of the dust in the HD 202628 disk
until the grains are ground down to a radiative blowout radius of $\sim$1
$\mu$m and are expelled from the system.  HD 202628 is younger than the Sun
with an enhanced X-ray luminosity of $L_X/L_{X_\odot} = 10^{28.41}/10^{27.35} =
11.5$.  Using Equation 3 from P09 we estimate a stellar wind with
$\sim26\times$ the solar wind mass loss rate, assuming that the radius of HD
202628 is approximately equal to that of the Sun.  Using Equations A10 and A16
in P09 we estimate the factor $P_{CPR}\approx9.7$ that the enhanced
stellar wind contributes to shortening the Poynting-Robertson drag timescale.
Finally, using Equation A23 in P09, with $L_{dust}/L_{star} =
1.4\times10^{-4}$, $D_{ring} = 158$ AU, $Q_{coll} = 1$, and assuming the mass
and luminosity of HD 202628 matches the Sun, we estimate that the grain-grain
collision timescale is $\sim$100 times shorter than that for the grains to
spiral inwards towards the star under the combined effects of stellar wind and
Poynting-Robertson drag.  Equation A23 in P09 is independent of factors such as
the grain size and density; it assumes a relative disk annulus of 0.1 times the
disk diameter, whereas the disk for HD 202628 is four times as thick in annular
diameter.  Therefore, this timescale ratio is only a lower bound, but we can
definitively conclude that grain-grain collisions dominate the dust grain
evolution of this disk and that the dust grains originate from the collisions
of larger parent bodies within the same annulus (e.g. Kuiper-belt object
analogues).

\subsection{Comparisons to Other Ring-Like Disks}

There are a variety of ring-shaped debris disks that have been imaged in
scattered light. The widths of the rings, the sharpness of their inner and
outer edges, their ellipticities (or lack thereof) vary widely.  Here we
summarize them to provide some context for the HD 202628 ring.  Note that we
omit extended disks with central clearings (e.g., $\beta$ Pic).

Of all the other disks, HD 202628's seems most like Fomalhaut's.  Fomalhaut's
ring has a sharp inner edge with similar surface brightness profiles of
$r^{10.9}$ along the inner edge and an outer falloff of $r^{-4.6}$ (Kalas et
al.\  2005).  The Fomalhaut ring is smaller than HD 202628's (113 -- 158 AU;
$\Delta r = 25$ AU; $\Delta r / r = 0.2$; unless otherwise noted, ring sizes
given here represent the $>50$\% of maximum surface brightness zones), and it
has a lower eccentricity ($e$ = 0.11 versus 0.18) with a smaller star/ring
offset (15.3 AU vs. 28.4 AU).  The Fomalhaut $L_{dust} / L_{star} = 8 \times
10^{-5}$ is about half that of HD 202628, and the star is younger ($\sim200$
Myr). 

HD 181327 (F5V, Schneider et al.\ 2006; Lebreton et al.\ 2012) has a ring-shaped
disk extending from 80 -- 100 AU ($\Delta r = 20$ AU; $\Delta r / r = 0.2$),
though a faint halo of dust can be seen out to $\sim455$ AU.  This is a much
brighter and more massive disk ($L_{dust} / L_{star} = 2 \times 10^{-3}$) than
HD 202628's or Fomalhaut's.  Curiously, its surface brightness profile is very
similar to Fomalhaut's and HD 202628's, with the inner edge rising as $r^{10}$
and the outer disk falling as $r^{-4.7}$ (Lebreton et al.\ 2012).  There is no
reported eccentricity. 

HR 4796 (A0V) is quite young (8 Myr), and it has one of the brightest debris
disks ($L_{dust} / L_{star} = 5 \times 10^{-3}$).  Its ring-shaped disk is
small and narrow ($r_{in} \approx 65$ AU, $r_{out} \approx$ 91 AU;$\Delta r = 25$
AU; $\Delta r / r = 0.3$).  Both its inner and outer edges are sharp; the cause
of the outer edge truncation has yet to be satisfactorily explained, but it has
been demonstrated that it cannot be due to a known, nearby stellar companion
(Th{\'e}bault, Marzani, \& Augereau 2010).  Recent adaptive optics imaging
(Thalmann et al.\  2011) verified a 1.2 AU offset along the apparent major axis
seen by Schneider et al.\ (2009), but a $\sim$5 AU offset along the line of
sight was also seen.  This suggests an eccentricity of $e \approx 0.07$. 

HD 207129 (G0V; Krist et al.\ 2010) has a faint ($V$ = 23.7 mag arcsec$^{-2}$),
ring-shaped disk extending over 148 -- 178 AU ($\Delta r = 30$ AU; $\Delta r /
r \approx 0.2$; note that these are the visible, not 50\% brightness, extents).
There is no indication of an offset of the ring or any significant azimuthal
density variations. Due to low signal, reliable surface brightness profiles
could not be obtained.  It has little or no forward scattering.

HD 107146 (G2V; Ardila et al.\ 2004) has a low-inclination disk with an inner
clearing at $r \approx 60$ AU.  The surface brightness increases gradually out
to 130 AU and then falls just as gradually further out. The 50\% surface
brightness radii are 87 AU and 168 AU ($\Delta r = 81$ AU; $\Delta r / r =
0.6$).  There is no observed eccentricity.  

HD 92945 (K1V; Golimowski et al.\ 2011) has a disk that extends from $r =$ 43
-- 140 AU.  The inner clearing is not fully defined in the $HST$ images, but
there is a density enhancement along the inner edge at 40 -- 70 AU.  The outer
edge also has enhanced density up to 100 AU, beyond which the surface
brightness falls off dramatically ($\sim r^{-10}$). This may indicate that
there are blended inner and outer rings.  The disk has no notable eccentricity.   

Table 1 summarizes some of the important properties of these disks.  There
seems little in common among them, except that they are, by selection, rings.
The inner radii (based on $>50$\% intensity) vary considerably, from 40 -- 150
AU.  The sharp inner edges of some (HD 202628, Fomalhaut, HR 4796, and perhaps
HD 181327) are likely defined by the orbital radii and masses of planets that
are truncating them.  The existence of planets is further bolstered by the
eccentricities seen in at least three of the rings (HD 202628, Fomalhaut, and
HR 4796). In contrast, HD 107146 has a gentle radial brightness profile without
a sharp inner edge.  Ertel et al.  (2011) and Hughes et al.  (2011) suggest
that this favors a morphology induced by a collisional cascade of planetesimals
unaltered by planetary perturbation.  The sharpness of the inner edge of HD
92945 is uncertain due to subtraction residuals near the star (Golimowski et
al. 2011) and of HD 207129 due to poor signal (Krist et al.  2010).  In two
cases, HR 4796 and HD 92945, the outer disk extents are sharply truncated,
perhaps due to unseen companions (note that the outer radii of both are smaller
than the inner radius of HD 202628).  All of the other disks have gradual
brightness falloffs with increasing radius, characteristic of grains being
blown out by radiation pressure and winds, and, likely in some cases, bumped
into higher eccentricites by planets (HD 202628 is uncertain due to poor
definition).  

\begin{deluxetable}{lcccccl}
\rotate
\label{tab:disks}
\tablewidth{7in}
\tablecaption{Disks with Ring-like Architecture}
\tablehead{
\colhead{Star} & \colhead{Spectral} & \colhead{$L_{\rm dust}/L_{\star}$} & \multicolumn{3}{c}{Ring Width\tablenotemark{1}} & \colhead{} \\
\cline{4-6} \\
\colhead{Name} & \colhead{Type} & \colhead{$\times10^{-4}$} & \colhead{$\Delta r$ (AU)} & \colhead{$\Delta r /r$} & \colhead{$e$} & \colhead{Comments} }
\startdata
HD 202628 & G2V & 1.4 & 150--220 & 0.4       & 0.18 &  \\
HD 207129 & G0V & 1.4 & 148--178 & 0.2       & none & visible range, not half-brightness\\
Fomalhaut & A3V & 0.8 & 113--158 & 0.2       & 0.11 &  \\
HD 107146 & G2V & 8   &  87--168 & 0.6       & none &  \\
HD 181327 & F5V & 20  &  80--100 & 0.2       & none\tablenotemark{2} & halo seen out to 455 AU \\
HR 4796   & A0V & 50  &  65--91  & 0.3       & 0.07 &  \\
HD 92945  & K1V & 8   &  40--63,117\tablenotemark{3} &  0.4,1.0\tablenotemark{3} & none & two blended rings? \\
\enddata
\tablenotetext{1}{Unless otherwise noted, ring widths correspond to the region with surface brightness 
$>$ 50\% of maximum.}
\tablenotetext{2}{No eccentricity was reported for HD 181327 based on $HST$/NICMOS
observations. However, higher resolution $HST$/STIS observations have been
obtained in GO Program 12228 (PI=Schneider) that are likely to provide a better
constraint on eccentricity, though results from that study were not available at 
the time of writing.}
\tablenotetext{3}{For HD 92945, the first outer radius given is that of the inner ring, the second is of the
outer ring; the width ratios are, respectively, for the inner ring only and the combined rings.}
\tablecomments{For a complete scattered light debris disk list, see Golimowksi et al.\ 2011.}
\end{deluxetable}

\section{Conclusions}

HD 202628 is one of only three stars, including Fomalhaut and HR 4796, with
eccentric, ring-shaped debris disks with sharp inner edges, all signs of
planetary tidal interactions. It has the largest and most eccentric ring of the
three.  It is also the only solar-type star and is much older ($\sim2$ Gyr)
than the other two. Based on the previous modeling of the Fomalhaut ring
(Quillen 2006; Chiang et al. 2009), the perturbing planet in the HD 202628
system is probably close to, and inside of, the $\sim158$ AU inner edge. This
shows that planets at very large orbital radii are present around not only more
massive stars like Fomalhaut, which had presumably more massive accretion
disks. This may be evidence that planets may form closer to the stars and get
scattered to large distances by other, more massive planets.  

We hope to obtain deeper exposures of HD 202628 with $HST$ that would show this
faint disk with better definition.  These would allow more accurate
characterizations of the inner edge and the outer extent of the disk.  These
will allow the use of modeling, as in Chiang et al. 2009, to more precisely
constrain the location and mass of the perturbing planet.

\section{Appendix}

Images of the other nine stars in our program (Table 2) revealed no signs of
circumstellar nebulosity.  The quality of the PSF subtraction for each was
similar to that obtained for HD 202628.  As an example, Figure 7 shows one of
the candidates, HD 145229, before and after PSF subtraction.

To determine detection limits in these images, we followed the procedure used
by Krist et al.\ (2010). They determined that a robust visual detection limit
for an extended source (specifically, a 1\arcsec $\times$ 1\arcsec uniform
square) is 1.2 times the local root-mean-square (RMS) noise level.  In each of
our roll-subtracted images, the RMS of all pixels at each radius was computed,
producing a radial noise profile.  Visually-identified bad (hot) pixels and
background objects were omitted.  The noise curves were multiplied by 1.2 and
converted to surface brightness in V magnitudes per square arcsecond (assuming
neutral scattering).  The conversion used the predicted count rate for each
target provided by the STIS exposure time calculator based on the star's V
magnitude and spectral type.  The results are shown in Figure 8.  Rebinning by
2 $\times$ 2 pixels (to 0\farcs 1 pixels) shifts these curves downwards by
$\sim0.5$ mag. The curves for the brightest stars are dominated by subtraction
residuals at all radii, while for most targets the detection limit converges to
24 mag arcsec$^{-2}$ at larger angles due to detector noise.

HD 202628 does not have any characteristics that distinguish it from the other
stars, so there is no clear explanation why more disks in our sample were not
detected.  As noted before, a face-on disk would be self-subtracted out by the
roll subtraction algorithm.  However, the chances of any given disk being
viewed face-on is low, and if one were, it would be difficult to detect
regardless of the subtraction method due to the low integrated column depth
(the most face-on disk seen in scattered light so far is HD 107146 (Ardila et
al.  2004), with an inclination of $\sim$25\deg\ from pole-on and an $L_{dust}
/ L_{star}$ = $1.2 \times 10^{-3}$, which is about an order of magnitude
brighter than our targets).  Other causes are discussed in Krist et al.\
(2010). 

Two of our targets are notable. HD 38858 has a {\it Spitzer}-resolved (70
$\mu$m) disk (Bryden et al., in preparation) with a radius of 9\arcsec (135
AU), inclination of 48\deg, and the apparent major axis along PA = 56\deg\ (our
observations of this target were specified to place the disk major axis
perpendicular to the occulting wedge).  The star also has a
radial-velocity-detected planet (HD 38858 b; Mayor et al.\ 2011) with a period
of 407.1 days, semi-major axis = 1.04 AU, and $m$ sin $i$ = 30.6 $M_{earth}$.
HD 45184 also has an R-V planet (HD 45184 b; Mayor et al.\ 2011) with a period
of 5.9 days, semi-major axis = 0.06 AU, and $m$ sin $i$ = 12.7 $M_{earth}$.  

By themselves, these $HST$ detection limits place weak constraints on the dust
properties of the disks.  However, when combined with resolved emission images
of these disks obtained with {\it Spitzer} or {\it Herschel}, they provide very
strong constraints on albedo and grain size distributions.  We will be
reporting these parameters for the non-detections from this and other $HST$
programs. 

\begin{figure}\label{hd145229}
\epsscale{1.0}
\plotone{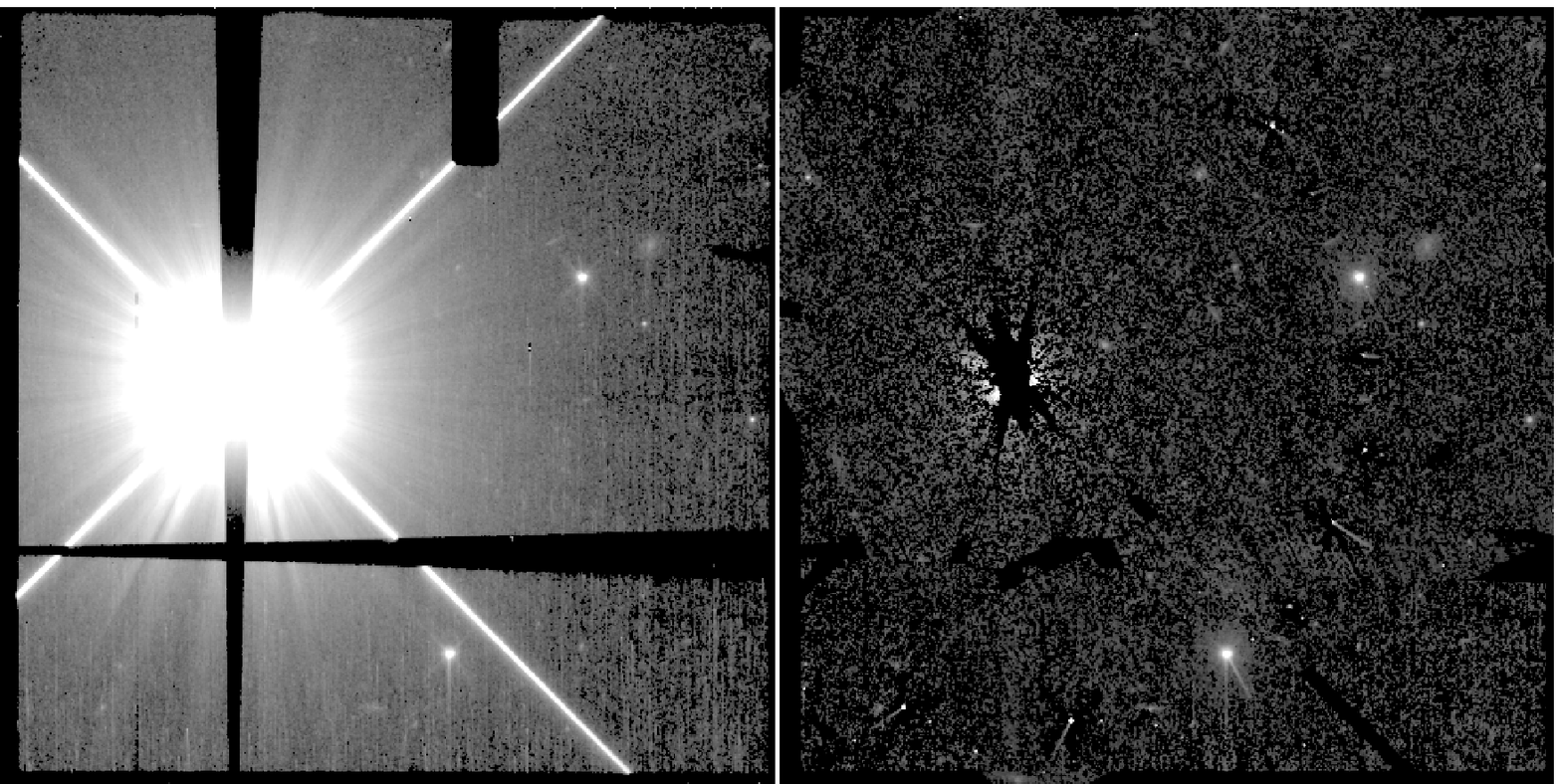}
\caption{STIS coronagraphic images of HD 145229, both displayed with identical 
quarter-root intensity stretches. The images are 52\arcsec on a side. (LEFT) PSF unsubtracted image from the first
orientation. The occulting mask shadows and diffraction spikes from the star
are evident. (RIGHT) PSF-subtracted, combined images from the two orientations produced
by the iterative roll subtraction algorithm. }
\epsscale{1.0}
\end{figure}

\begin{figure}\label{nondetections}
\epsscale{1.0}
\plotone{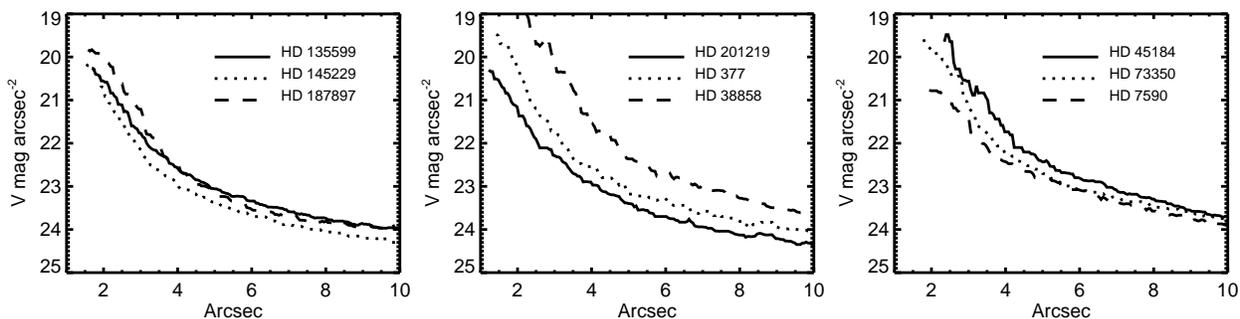}
\caption{Lower surface brightness limits for the reliable visual detection of a 1\arcsec $\times$ 1\arcsec square
around targets for which no disks were seen.}
\epsscale{1.0}
\end{figure}

\begin{deluxetable}{lcccrccl}
\rotate
\label{tab:targs}
\tablecolumns{8}
\tablewidth{0pc}
\tablecaption{Debris Disk Non-Detections in {\it HST} Program 12291}
\tablehead{
\colhead{} & \colhead{Spectral} & \colhead{$V$} & \colhead{Distance} &
\colhead{Observation} & \colhead{$L_{\rm dust}/L_{\star}$} & \colhead{{\it Spitzer}} & \colhead{} \\
\colhead{Target} & \colhead{Type} & \colhead{(mag)} & \colhead{(pc)} &
\colhead{Date} & \colhead{$\times10^{-4}$} & \colhead{Reference\tablenotemark{1}} & \colhead{Comments}}
\startdata
HD    377 & G2 & 7.6 &  39 & 22 Jul 2011 & 4 & H08 & NICMOS non-detection\\
HD   7590 & G0 & 6.6 &  23 & 25 Oct 2010 & 3 & P09 & age 500 Myrs\\
HD  38858 & G4 & 6.0 &  15 & 24 Dec 2011 & 1 & B09 & disk resolved by {\it Spitzer}, has planet\\
HD  45184 & G2 & 6.4 &  22 &  3 Oct 2011 & 1 & K10 & has planet\\
HD  73350 & G5 & 6.7 &  24 & 24 Nov 2011 & 1 & P09 & age 500 Myrs\\
HD 135599 & K0 & 6.9 &  16 &  2 May 2011 & 1 & P09 & age 200 Myrs\\
HD 145229 & G0 & 7.5 &  35 &  3 Jun 2011 & 1 & H08 & NICMOS non-detection\\
HD 187897 & G5 & 7.1 &  35 & 10 Oct 2010 & 1 & H08 & NICMOS non-detection\\
HD 201219 & G5 & 8.0 &  38 &  2 Nov 2010 & 1 & H08 & NICMOS non-detection\\ 
\enddata
\tablenotetext{1}{B09: Bryden et al.\ (2009), H08: Hillenbrand et al.\ (2008), 
K10: Koerner et al.\ (2010), P09: Plavchan et al.\ (2009)} 
\end{deluxetable}

\acknowledgments{ 
Based on observations made with the NASA/ESA {\it Hubble Space Telescope},
obtained from the data archive at the Space Telescope Science Institute. STScI
is operated by the Association of Universities for Research in Astronomy, Inc.
under NASA contract NAS 5-26555.  The {\it Spitzer Space Telescope} is operated
by the Jet Propulsion Laboratory, California Institute of Technology, under
NASA contract 1407.  This work was funded by NASA via a {\it Hubble Space
Telescope} General Observer Grant (program 12291) to the Jet Propulsion
Laboratory, California Institute of Technology and the {\it Spitzer} Project
Science Office at JPL. 
}

\end{document}